# Digitality as a "longue durée" historical phenomenon

Salvatore Spina
Department of Humanities – University of Catania
salvatore.spina@unict.it

## Abstract

The digital age introduced the Digital Ecological Niche (DEN), revolutionizing human interactions. The advent of Digital History (DHy) has marked a methodological shift in historical studies, tracing its roots to Babbage and Lovelace's 19th-century work on "coding" as a foundational communication process, fostering a new interaction paradigm between humans and machines, termed "person2persons2machines." This evolution, through digitization and informatization, builds upon ancient coding practices but was significantly advanced by Babbage and Lovelace's contributions to mathematical linguistic systems, laying the groundwork for Computer Science. This field, central to 20th-century mainframe interaction through programming languages and formalization, situates Digital History within a broader historical context. Here, coding and mathematical methodologies empower historians with advanced technologies for historical data preservation and analysis. Nonetheless, the extent to which computation and Turing machines can fully understand and interpret history remains a subject of debate.

## Keyword

Encoding, Digitalization, Digital Ecological Niche (DEN), Punched Cards, Abstraction, Babbage, Lovelace

## 1. The DEN, the coding

The most recent "mid-term" period has been characterized by the advent of digital transformation, specifically the establishment of the Digital Ecological Niche (DEN) (Spina 2023c; 2023d), a sociocultural habitat formed through a triad of interconnected nodes, in accordance with the "person2persons2machines" dynamic.

Differing from the ecological niches (Godfrey-Smith 2000; Iannucci 2020; Laland, Odling-Smee, and Feldman 2000b; 2000a; 2005; Laland and O'Brien 2011; Latina 2013; Odling-Smee and Turner 2011; Steen 2000; Uller and Helanterä 2019) composed of all living organisms and their natural habitat, the DEN operates as an advancement of the communication system. Within this system, the agents consist of not only humans but also machines, whose presence prompts a new process of symbol textual encoding in a format readable by machines. The transition to digital technology, in essence, refers to the "computerized digitization of the «original digitization»", which pertains to the initial process of character encoding through the development of vowels and consonants, utilized by individuals to convey, elucidate, and symbolize.

The clear-cut and incontrovertible postulate has defined the means of communication utilized amongst humans. In light of this idea, Claude Elwood Shannon (Shannon 1948; 1993; Nahin 2017; Shannon and Weaver 1998) and Warren Weaver (Nahin 2017; Shannon 1993; Shannon and Weaver 1998) developed their theories[1] and studies, leading us to contemplate "digitalization" as a process that stems from the human intellect.

The linguistic studies of de Saussure (de Saussure 1879; 2014) and the Russian formalists[2] are credited with laying the foundation for the modern Theory of Communication, as they sought to systematize the formal aspects of communication. In light of the subject matter at hand, it is worth noting that the most paramount notion is that of Shannon, who in the year 1945 embarked on a quest to ascertain a practical measure by which information can be "quantified. This system would allow for the exchange of information between two calculators. Consequently, Shannon perceived communication through a mathematical lens, where data could be streamlined to eliminate duplication and thereby eliminate superfluous and non-quantifiable information. Shannon's endeavours can be aptly identified as a mathematical interpretation of encoding, leading to digitization, entailing the development of a discrete and actionable message.

In a similar vein, Warren Weaver expounded in his *Memorandum* to Norbert Wiener (Wiener 1953), the founder of cybernetics, the prospect of implementing computer systems for automatic language translation. The foundation of this concept is rooted in the notion that language serves as a system of encoding.

In the present day, the most recent digital revolution, integrating machines as active participants in the interplay structure, inevitably requires a deeper level of abstraction in the communication system. As a result, it is imperative to establish a novel structure and relocate all modes of discourse to the DEN. In the wake of Turing's influence, as predicted, machines have developed into agents that produce knowledge.

In light of the digital transformation being indicative of the 'mid-term', it is equally imperative to acknowledge that this process must be regarded as a long-term historical phenomenon (Braudel 2003; 2015; Braudel and Salsano 1974), which has only been acknowledged as a significant historical issue in recent times. In spite of this, within the current timeframe, contemplation has dissociated the expression from its significance and denotation, resulting in a growing cognizance among scholars that digitalization is a deceitful guise, misleading one's perception instead of advancing it.

In light of the ongoing discourse, albeit veiled, within the scope of a humanistic methodology that continues to reveal a dearth of comprehension regarding the concept of digitalization, its "long-term" remains imperceptible to many and is entirely overlooked in departmental and interdepartmental undertakings. The intellectuals, who consistently recycle theoretical

---

[1] The "Information Theory", the former, and the possibility of automatically translating from one language to another, the latter.

[2] We can read about Viktor Borisovič Šklovskij, Vladimir Jakovlevič Propp, Roman Jakobson, Boris Michajlovič Ėjchenbaum and the Linguistic Center of Moscow in the book written by Todorov and Bravo (2003)

standpoints that do not aim to amalgamate the "qualitative" and "quantitative", "event-based" and "universal", satisfy themselves by discovering remedies through the addition of suffixes and prefixes to the term «History».

As an illustration, the term «hyperHistory» (Floridi 2012; 2016; 2020), in an endeavour to amplify its significance, to additionally connect it to the era of knowledge and intangible assets, while disregarding the fact that such classifications hold no substance.

Taking into consideration the prefix "hyper" in its linguistic sense, signifying "going beyond", it can be concluded that History, while not transcending itself, does not have the intention to do so —a rational assertion.

The phenomenon of "communication acceleration" and "service enhancement" have no direct correlation with History, except when the latter chooses to focus its attention on elucidating that historical and historiographical context. By altering the initial point of reference to the moment of conceptualization of "service", it initiates the mechanism by which digital communication bears no distinction from the offerings of previously existing communication frameworks.

In considering "hyperhistory" as a process of transcending boundaries, we would simply engage a whimsical mechanism that would prompt us to contemplate the realm of potentialities, thereby converting Information Technology into a medium of communication that would metamorphose History into a Sci-Fi tale. In spite of Nadel's conviction that History is endowed with the capacity to prognosticate the future (Nadel 1964), its essence remains unchanged, rendering hyperhistory a mere moment in its bygone existence, designated solely as History, a term that cannot be modified by prefixes or suffixes under any circumstances.

The digital revolution has indeed expedited the progression of History, characterized as a meticulous procedure in which a concluded and perfected action swiftly transforms into an archival "record," from which it promptly reemerges as a "source". Our interlude is unequivocally defined by an accelerated decrease in the timeframe needed to examine a source and its final dissemination as a research output (Spina 2023c).

However, it is indisputable that this rapid advancement falls within the realm of History and never surpasses it. Furthermore, the notion of "presentism" (Hartog 2015) —the misconception of the steadfastness of the current moment— does not alter the concept of History towards a state of hyperactivity. The expeditious tempo of the *hic et nunc* in the digital domain does not disassociate present actions from their past occurrences, as they persist from their inception.

The succinct nature of interactions among the representatives of the DEN is in line with the overarching "long-term phenomenon" of digitalization, which is —and must expressly revert to being— the process of "encoding," as previously stated. The genesis of a code, emerging from the primary endeavour that culminated in the formation of the alphabet (Duranti 2005; Haarmann 2021; Johansson 2023), facilitates the dissemination of knowledge between two parties/entities.

As we are aware, the emergence of human characteristics was a gradual process, and Science is unable to definitively distinguish between primates and our forebears. Likewise, a definitive moment cannot be ascertained regarding the mutual understanding among humans on the nomenclature adopted for giving names to physical objects, geographical sites, and metaphysical entities. There has never been a linguistic Big Bang, as theorized by Noam Chomsky (Chomsky 1975). Notwithstanding, regardless of the route taken, be it through biology, physiology, or cognition, it is indisputable that, through a series of events beyond our complete comprehension, primitive humans' grunts and vocalizations evolved into diverse utterances (terms) utilized to depict varying circumstances.

The set of letters (vowels and consonants) that has materialized as a result of this undertaking is the outcome of said act of codification, thereby constituting a framework for encoding, which signifies the commencement of the «information digitization process». The primary goal of codification is to establish a limited set of symbols capable of generating an unlimited range of expressions, with the utmost purpose of documenting humanity's memories and history (Dunbar 1998).

Digital transformation remains a long-term phenomenon, currently marked by a progression towards digital transmission of information. In this regard, a novel entity, the Turing Machine, coexists alongside human agents. The latter, devised as a medium for trade, is implemented at the discretion of the former and its continuous improvement endows it with the ability to generate data autonomously and function as an integral component of the network, assuming the role of a node.

With that being said, it must be emphasized that reflection does not absolve, nor does it simplify, the recognition of the emergence of the DEN from the material realm. However, the upgrading procedure, as a result of the digital transformation, the DEN is assimilating the physical world, intending to encode it to generate its digital counterpart, encompass not only human creations (artefacts and records), in both corporeal and non-corporeal forms, but also striving towards the substitution of analogue services with entirely digital ones. The establishment of the DEN is rooted in the imperative of an anthropological transformation that designates the individual as *Homo-Loggatus*, an integral component of the DEN structure, achieved by an act that generates a digital iteration of humanity, retaining solely its informational components[3]. The aforementioned procedure compels historians to engage in introspection, directing their focus towards themselves rather than History as a whole. This necessitates an evaluation that propels their field towards the realm of Digital History (DHy).

---

[3] The concept of "Homo-Loggatus" refers to humans who have acquired an active digital identity in the digital world. This concept is based on the notion that through participation and interaction with digital technologies and computer infrastructures, humans become an integral part of the "digital ecological niche." Essentially, becoming a "Homo-Loggatus" implies that the individual has connected or "logged in" to the digital network through various platforms and technological tools. This connection provides the individual with a digital identity, which may include personal data, preferences, online activities, and more. This digital identity thus becomes a significant part of the individual's life and experience, influencing their interaction with the digital world and their role within it. The "Homo-Loggatus" actively participates in digital culture, contributing content, interactions, and online transactions. Additionally, this digital identity can influence access to digital services, the creation of online communities, and even economic opportunities. The "Homo-Loggatus" represents a new type of humans who have adapted and integrated into the digital society, assuming a significant identity and role within the context of the "digital ecological niche," as stated by (Spina 2023c; 2023d)

## 2. Digital History as History of Digitality?

The very foundation of the DEN can be traced back to the process of encoding, which establishes a linguistic framework for seamless communication between human beings, groups, and machines. In this regard, Computer Science (Informatics) serves as the quintessential "linguistic unifier" among the Sciences, offering a universal code to streamline the intricacies of multidisciplinarity and interdisciplinarity.

It is of utmost importance to recognize that this conjecture inherently categorizes historians as "informatic agents," as posited by Le Roy Ladurie (Le Roy Ladurie 1968). In essence, they must promptly adapt to coding and Information Technology in their professional endeavours[4]. It is crucial for historians to acquire a skillset that will guide them to consider ICT and artificial intelligence platforms as resources for a groundbreaking methodology that transforms the discipline of historical science. It is imperative to recognize that the Turing Machine offers a multitude of advantages over the traditional approach, particularly in the realms of calculation, storage, processing, and retrieval of information. It goes without saying, furthermore, that digital historians are not such because they replace paper, pen, and typewriter with a computer, remaining, in this perspective, within the realm of traditional methodology. Conversely, the discipline of Computer Science will guide historians to perceive Digital History as an essential chronicle and historiography of the Digital Era Network, representing an advancement in communication networks. In the interim, the arrangement and ongoing development of said system as a collaborative setting has altered human behaviour within it, establishing itself as a realm that warrants a narrative of its own history.

Thus, the configuration of DHy entails a twofold approach, requiring historians to navigate their investigations both beyond the DEN in a conventional manner, and within it through digital means, in order to effectively present the evolving narrative of History.

To categorize "Digital History" exclusively based on the methodological dynamic between humans and machines is overly reductionist. The matter at hand necessitates consideration on two fronts: firstly, does DHy serve as a linguistic-semantic methodology utilized by historians, namely a research process that solely relies on the utilization of Information Technology, thereby relegating historians to the role of "those who can give meaning" and effectively rendering computers as replacements for traditional tools such as paper, pen, and typewriter? Perchance (2) revisiting the concept of "digitalization" to its essence of "coding," DHy entails a computational procedure that empowers the Turing Machine to actualize insights on previous occurrences —consequently directing contemplation towards the query "Can a Machine document the Past?

Examining the primary scenario, the digital conversion is outlined as a progression that, on the one hand, has augmented historians' work surface with a plethora of fresh sources (digital and digitized), whose organization inherently demands the utilization of ITC for retention and assessment. Conversely, it transforms digitalization into a mere process of accumulation, thereby obliging archives to accommodate the historians' demand for online source accessibility through a procedure restricted to the photographic reproduction of paper records.

In this context, the second scenario, which fundamentally questions the role of historians, necessitates the examination of digitality and its complexities as a historical predicament. This, in turn, compels scholars to employ Information Technology as a tool for analysis. Conversely, in light of the matter, it is crucial to review the feasibility of a semantic introspection arising from an artificial intelligence platform, which will be a requisite for overseeing the extensive Big Data of History. The digitization of our cultural heritage necessitates a well-crafted encoding method that can convert archival materials into computerized information complexes. This will grant the Turing Machine the capacity to carry out computational procedures and generate fresh historical insights.

In light of this moment of reflection, it is imperative to recognize digitization as a long-term phenomenon that has brought about a revolutionary change in the daily lives of individuals in the digitalized society.

## 3. At the origins of the coding's long-term phenomenon

The digitalization process experienced a significant acceleration and oscillation between the 1930s and 1950s of the 20th century, ultimately becoming a pivotal component of the digital society, culture, and economy. The physical world has embraced digitalization, which, having originated from it, has now pervaded it entirely, assimilating it and transforming its equilibrium to suit the interconnectivity of "data" (Big Data). The latter, or "unphysical items," have become an abundant asset in the contemporary market, ultimately surpassing the discourse surrounding the potential of globalization. This has manifested in the emergence of formidable entities that possess the ability to penetrate once-impenetrable exchange realms.

The current direction of our progress, focused solely on "presentism" and utilizing the term in place of "future," has resulted in a significant disconnect between the original intention of the Turing Machine, its programmers, and its current state of wrong utilization.

It is imperative to engage in rebalancing to explicate the source of integrating the third entity of the DEN and, consequently, the divisive force driving the oscillation in the phenomenon.

Our rationale directs us towards the Victorian era, prompting us to examine the concepts of Charles Babbage and the musings of the mathematician Ada Byron Lovelace.

---

[4] «Grand mangeur d'informations, l'ordinateur-historiographe s'accommode du reste des problématiques, voire des idéologies les plus diverses. L'une des premières études d'«histoire-machine» parue voici quelque temps dans la revue Annales était l'œuvre d'un chercheur soviétique qui voulait établir le taux d'exploitation des paysans russes par les grands propriétaires d'autrefois: c'était du Marx ou du Lénine tout pur, mais accommodé à l'électronique. Dans un ordre d'idée un peu différent, aux États-Unis, les nouveaux historiens radicaux, comme Lockridge, qui tentent de réévaluer la révolution de 1776 et qui veulent lui trouver un contenu révolutionnaire, voire castriste, effectuent cette recherche avec la technologie la plus «sophistiquée»: dépouillant, au moyen d'ordinateurs, les centaines de milliers de chiffres contenus dans les documents fiscaux des treize colonies, ils tentent de montrer que les soulèvements de la guerre d'Indépendance procédaient d'un état de crise sociale: les petits fermiers, victimes de cette dépression, paupérisés par le morcellement de leurs terres, polarisèrent leur ressentiment contre les maîtres britanniques».

At that pivotal moment in History, the world was starting to mechanize and break down actions into smaller components. In the antecedent century, Gottfried Wilhelm Leibniz had previously condensed the alphabet into binary code, transforming each word into a sequence composed solely of two numerical components, «0» and «1». This procedure facilitates more efficient supervision of regulatory, computational, and operational endeavours. In the succeeding century, Harry Ford would adopt this principle and include it in his concept of the "assembly line".

Blaise Pascal conceived the notion to devise his calculating apparatus, aptly named the Pascaline, with the sole intention of aiding his father, Étienne Pascal (1588-1651), in the administration of his office as Superintendent of Finances in Rouen. Although the philosopher's intuition is decisive, «the much-admired machine of Pascal —according to Ada Lovelace— is [...] simply an object of curiosity, which, whilst it displays the powerful intellect of its inventor, is yet of little utility in itself. Its powers extended no further than the execution of the first four operations of arithmetic, and indeed were in reality confined to that of the first two, since multiplication and division were the result of a series of additions and subtractions. The chief drawback hitherto on most of such machines is, that they require the continual intervention of a human agent to regulate their movements, and thence arises a source of errors; so that, if their use has not become general for large numerical calculations, it is because they have not in fact resolved the double problem which the question presents, that of correctness in the results, united with economy of time» (Lovelace 1843).

The implementation of digitalization has, albeit unintentionally and inadvertently, permanently altered the linguistic framework. As of that point, the dialogue among human beings could be efficiently condensed into minimal numerical proportions, thus paving the way for the reduction of words and meanings into "second-order abstractions" elements, as declared by Frege (Bianchi 2003; Frege 1980; 2007; 2021; 2023), Hilbert (Corry 2013; Hilbert 1893; 1972; 2012; 2015; 2021; Hilbert and Cohn-Vossen 1999; Lotman 1984; Roselló 2019), and Gödel (Gödel 1940; 2012; Rodriguez-Consuegra 1995; Whitehead, Russell, and Gödel 1986). The crucial theoretical concepts of the three philosophers paved the way for a new understanding of "abstraction processes"[5] in the emerging century, which ultimately led to advancements in software and programming (Liskov and Guttag 1986). From thenceforth, our digitalization undertaking gave rise to the prospect of acknowledging a new entity within the framework of interdependencies: the machine.

Amidst the 1800s, many academics dedicated their efforts towards devising computational contrivances. The world was undergoing a phase that would revolutionize every invention hitherto utilized. A prime example of this is the replacement of carts and carriages with railways and the subsequent transition from sailing vessels to steamships. Every mechanical devising was open to enhancement.

The domain of mathematical computation may also be enhanced with the assistance of technological tools, prompting many mathematicians to devise computational devices. Within the group, we encounter a figure resembling Charles Babbage, who, during the 1820s and 1830s, endeavoured to create the "analytical engine," a programmable mechanism capable of carrying out a series of instructions through a designated «program».

The aforementioned invention lacks any novelty that surpasses the functionality of basic calculators, specifically the arithmometer devised by Charles Xavier Thomas de Colmar (Lazard and Mounier-Kuhn 2016). Notwithstanding, the primary purpose of the analytical engine was to obviate the laborious duty of perusing the outcome of each operation, transcribing it by hand onto paper, and reinserting it into the apparatus for the ensuing operation, thus sparing time and eradicating inescapable transcription mistakes. The analytical engine has the capability to operate autonomously until the desired result is reached.

<span style="color:red">Babbage's "computer" was to be comprised of a "mill" and a "store."[6]</span> The former was to perform the four fundamental arithmetic operations, while the "store" was a set of columns of decimal wheels, akin to those of the <span style="color:red">"difference engine,"[7]</span> where numerical data to be processed, partial results, and final outcomes were recorded.

The issue lay in the 'software', namely the sequence of commands and instructions to be imparted to the machine, which also necessitated the function of "conditional branching," *i.e.*, the ability to change the sequence of operations automatically depending on the result attained at that moment.

The "computer" thus required a medium to store data for acquisition. The question arose of how to make the machine work and which tools to use to encode and simplify the data to be processed. The dataset had to be discrete and free from falsified data.

Babbage's perspective was not amiss. It was imperative to encode information in a machine-readable language. At this historical juncture, thoughts turned to digitalization as a coding process not aimed at transmitting information among human subjects (human-oriented digitalization) but between humans and machines (computer-oriented digitalization). With Babbage, we are witnessing the genesis of that oscillation which led to the conceptualization of "encoding" as an essential step in the realization of the digital world and the establishment of the interplay structure upon which it is based: "person2persons2machine."

[5] «Abstraction is a way to do decomposition productively by changing the level of detail to be considered. When we abstract from a problem we agree to ignore certain details in an effort to convert the original problem to a simpler one. We might, for example, abstract from the problem of writing a play to the problem of deciding how many acts it should have, or what its plot will be, or even the sense (but not the wording) of individual pieces of dialogue. After this has been done, the original problem (of writing all of the dialogue) remains, but it has been considerably simplified perhaps even to the point where it could be turned over to another or even several others. [...] The paradigm of abstracting and then decomposing is typical of the program design process: Decomposition is used to break software into components that can be combined to solve the original problem: abstractions assist in making a good choice of components. We alternate between the two processes until we have reduced the original problem to a set of problems we already know how to solve. [...] The process of abstraction ran be seen as an application of many-to-one mapping. It allows us to forget information and consequently to treat things that are different as if they were the same. We do this in the hope of simplifying our analysis by separating attributes that are relevant from those that are not».
[6] The mill itself used a collection of rotating barrels with pegs mounted on them for its internal state changes and management. The store had the capability to hold up to one thousand forty-digit numbers. It was programmed in a low-level language —similar to how assembly language is used today. The whole machine would have been powered by a steam engine (Babbage 2010).
[7] Unlike the "analytical" device, the "difference engine" worked by using polynomial functions, that is, a system that uses variables and coefficients, and fairy simple mathematical functions —addition, subtraction and multiplication. It did this by applying an algorithm called "divided differences", which consisted of a series of columns, each representing a number value with each column showing part of the result of the calculation. It was designed to make calculations with sixteen digits and six orders of magnitude —which is, a range between one and a million, or a thousand and a thousand million.

Luigi Menabrea seized upon Babbage's ideas and decided to describe the analytical engine in an article published in the «Swiss journal Bibliothèque universelle de Genève», titled *Sketch of the Analytical Engine by Mr. Charles Babbage* (Menabrea 1842). The text made its way to England, where Charles Wheatstone appreciated its contents, to the point of asking Ada Lovelace, who was thoroughly acquainted with Babbage's theories and projects, to translate it into English.

Ada agreed, but with no intention of writing a new article, she opted to merely append explanatory notes to Babbage's text, among which she added a method of calculating «Bernoulli Numbers,» which is seen as the first complete computer program, making Ada Lovelace the first computer programmer.

The final text, after a few months, was published in the journal «Scientific Memoirs, Selected from the Transactions of Foreign Academies of Science and Learned Societies», under the title *Sketch of the Analytical Engine Invented by Charles Babbage*, an essay that represents the theoretical synthesis between Leibnizian's thought and Pascal's mechanical devices. Ada's essay is the experimental idea of "software" as a linguistic system between humans and machines.

Lovelace's was certainly not a verbal system. The interplay required a physical medium on which to base the exchange, which Ada identified in the Jacquard loom: the backing system, which allowed, using punched cards, the loom to work automatically without the intervention of the human operator.

> «To simplify this manufacture, Jacquard devised the plan of connecting each group of threads that were to act together, with a distinct lever belonging exclusively to that group. All these levers terminate in rods, which are united together in one bundle, having usually the form of a parallelopiped with a rectangular base. The rods are cylindrical, and are separated from each other by small intervals. The process of raising the threads is thus resolved into that of moving these various lever-arms in the requisite order. To effect this, a rectangular sheet of pasteboard is taken, somewhat larger in size than a section of the bundle of lever-arms. If this sheet be applied to the base of the bundle, and an advancing motion be then communicated to the pasteboard, this latter will move with it all the rods of the bundle, and consequently the threads that are connected with each of them. But if the pasteboard, instead of being plain, were pierced with holes corresponding to the extremities of the levers which meet it, then, since each of the levers would pass through the pasteboard during the motion of the latter, they would all remain in their places. We thus see that it is easy so to determine the position of the holes in the pasteboard, that, at any given moment, there shall be a certain number of levers, and consequently of parcels of threads, raised, while the rest remain where they were» (Lovelace 1843).

Every machine responds to a fundamental principle: to enable humans to have more, in less time. Every mechanical invention must guarantee a "gain," whether it be in purely economic terms or in terms of the immediate resolution of a calculation and analysis problem (Spina 2022a; 2023c). While it is true that the Pascaline and the automatic loom respond to this need, it is even truer that it would have been Babbage's machine (Babbage 1989; Bromley 1991; Collier and MacLachlan 2000; Dasgupta 2014; Hyman 1985; Lovelace 1843; Swade 2002) which could have taken this reasoning to a more innovative level. But his creation failed; nevertheless, Ada Lovelace's reflection embodies the principle upon which today's Computer Science is based, up to the advent of "cloud" systems: the creation of a medium for the storage, organization, and transmission of data and documents (Floppy Disk, CD-ROM, DVD-ROM, etc.), for their subsequent computational analysis.

Certainly, distinctions are in order. If, for the loom, punched cards provided the correct instructions for executing the artwork in the fabric being worked on, the same punched cards (the "Operation Cards" and the "Cards of the Variables"), and their proper arrangement, would have worked differently on the English inventor's analytical engine: they would have, indeed, allowed it to perform addition, subtraction, multiplication, and division operations, and subsequently represent them in specific columns.

Hence, the historical root of encoding, as the systematization of a language system capable of enabling dialogue between humans and machines. We are at the first theoretical experience of the need to 'formalize', which will see, subsequently, in Jean-Claude Gardin a new "innovator". From that moment on, the meanings of the historical digitalization phenomenon changed, and the new concept led, in the subsequent centuries, to artificial intelligence tools and to all those computer technologies that are based on the codifying process to create machine-readable information. The codifying process will translate into a controversial mechanism of adaptation to the Digital Ecological Niche, in which linguistic expressions and words —the "social glue"— become relational links (edges) between "dimensions" (humans), which become nodes arranged on different servers, allowing connection to the Internet.

The "computer" thus required a medium to store data for acquisition. A dilemma emerged regarding the operation of the machine and which techniques to utilize for encoding and streamlining the data for processing. To ensure the dataset's integrity, it needed to be discrete and free from any falsified information.

Babbage's stance was not erroneous. It was imperative to encode information in a machine-readable language. During this pivotal moment in history, the focus shifted towards digitalization as a form of coding that was not intended for the exchange of information between humans (human-oriented) but rather between humans and computers (computer-oriented). Babbage's contributions mark the beginning of the oscillation that paved the way for the conceptualization of "encoding" as a fundamental element in the materialization of the digital sphere and the establishment of the interdependent structure known as "person2persons2machine."

In a demonstration of intellectual acumen, Mr. Luigi Menabrea embraced Babbage's theories and proceeded to outline the analytical engine in a piece published in the renowned "Swiss journal Bibliothèque universelle de Genève," entitled "Sketch of the Analytical Engine by Mr. *Charles Babbage* (Menabrea 1842). The document reached England, where Charles Wheatstone held great admiration for its contents and subsequently requested Ada Lovelace, who possessed extensive knowledge of Babbage's concepts and endeavours, to render it into English.

Ada has confirmed her agreement, although she has no plans to compose a new piece; rather, she has decided to add explanatory notes to Babbage's text. After a span of several months, the ultimate draft was released in the esteemed publication "Scientific Memoirs, Selected from the Transactions of Foreign Academies of Science and Learned Societies," bearing the title "Sketch of the Analytical Engine Invented by Charles Babbage." This composition serves as an amalgamation of Leibnizian ideology and the mechanical contrivances of Pascal. Ada's written piece delves into the experimental notion of "software" as a language system that bridges the gap between humans and technology.

Lovelace's was certainly not a verbal system. In order for the interaction to occur, a physical medium was necessary. Ada pinpointed the Jacquard loom as the ideal platform, as its backing system utilized punched cards to automate its functioning without requiring human input.

«To simplify this manufacture, Jacquard devised the plan of connecting each group of threads that were to act together, with a distinct lever belonging exclusively to that group. All these levers terminate in rods, which are united together in one bundle, having usually the form of a parallelopiped with a rectangular base. The rods are cylindrical, and are separated from each other by small intervals. The process of raising the threads is thus resolved into that of moving these various lever-arms in the requisite order. To effect this, a rectangular sheet of pasteboard is taken, somewhat larger in size than a section of the bundle of lever-arms. If this sheet be applied to the base of the bundle, and an advancing motion be then communicated to the pasteboard, this latter will move with it all the rods of the bundle, and consequently the threads that are connected with each of them. But if the pasteboard, instead of being plain, were pierced with holes corresponding to the extremities of the levers which meet it, then, since each of the levers would pass through the pasteboard during the motion of the latter, they would all remain in their places. We thus see that it is easy so to determine the position of the holes in the pasteboard, that, at any given moment, there shall be a certain number of levers, and consequently of parcels of threads, raised, while the rest remain where they were» (Lovelace 1843).

The cornerstone of all machinery is its ability to enhance human productivity, allowing for greater efficiency and output in a shorter time. Every mechanical invention must guarantee a "gain," whether it be in purely economic terms or in terms of the immediate resolution of a calculation and analysis problem (Spina 2022a; 2023c). Granted, the Pascaline and the automatic loom do meet this demand, yet it is without question that it was Babbage's machine (Babbage 1989; Bromley 1991; Collier and MacLachlan 2000; Dasgupta 2014; Hyman 1985; Lovelace 1843; Swade 2002) that could have advanced this thought process to a more cutting-edge level. Despite its ultimate failure, the work of Ada Lovelace remains a quintessential example of the fundamental tenet underlying modern Computer Science, even in the era of "cloud" technology: the development of a medium for the retention, arrangement, and dissemination of information and records (*e.g.* Floppy Disk, CD-ROM, DVD-ROM, etc.), for their subsequent computational analysis.

Certainly, distinctions are in order. In the circumstance that the punched cards were to offer the correct instructions for the execution of the design on the fabric, the aforementioned punched cards (referred to as the "Operation Cards" and the "Cards of the Variables") and their correct alignment, would have yielded varying outcomes on the analytical engine of the English inventor. In this given circumstance, they would have, indisputably, granted the capability to execute the functions of addition, subtraction, multiplication, and division, and subsequently illustrate them in assigned columns.

Consequently, the genesis of encoding can be attributed to the methodical categorization of a language system that facilitates discourse between humans and machines. We are presently encountering our initial theoretical encounter with the requirement to "formalize," a concept that will later be exemplified by Jean-Claude Gardin as a pioneering force. In the aftermath of this occurrence, the implications of the historical digitalization phenomenon were reshaped, and the ensuing concept brought about artificial intelligence instruments and all computer-based technologies that are predicated on the codifying process to produce machine-readable data. The codification process will yield a contentious adaptation mechanism for the Digital Ecological Niche, wherein linguistic expressions and words serve as relational links between "dimensions" (humans) organized as nodes on disparate servers, enabling Internet connectivity.

## 4. Coding as formalization

In the wake of the Lovelacean era, a century later, society found itself confronted with formidable machines, upon which hefty investments were bestowed by corporate entities and governmental bodies. Mainframes became coveted tools in every research field; even humanists took advantage, viewing digitality as a laboratory space capable of removing historians from the dimension of being "typographical scholars." Despite humanists historically holding a crucial role in language studies, the rise of digitalization has shifted their position and placed them on the sidelines. This was particularly evident during the development of a coding language for machine-readable texts. The conception of a predetermined fate, first proposed by Leibniz's Binary Code theory, must be confirmed through the essentiality, as emphasized by Ada Lovelace, of devising a simplistic language capable of mathematically translating our ideas.

The contributions of Lovelace were crucial in the development of specialized systems for communicating with mainframes, including Cobol, Basic, Ascii, Pascal, and Dos, by computer scientists like John Backus (Casalegno 2013; Slater 1989) and Grace Murray Hopper (Beyer 2012; Wallmark 2017).

The advent of calculators and computation in the 1930s marked a pivotal moment in Sciences, as it ushered in a period of rational and deliberate understanding. This era also brought about an awareness that the concept of "digitalization" had introduced novel "subjects" to the realm of relationships and Knowledge construction. Scholars were determined to gain specific competencies, and it was largely the responsibility of humanists to kickstart the contemplation process that could open up new opportunities. The advent of digitalization has positioned research in direct interaction with computers, thereby necessitating computation and continual encoding. As per Niklaus Wirth's observation (Wirth 2008), the cornerstone of Computer Science is comprised of data, its arrangements, and algorithms, and thus, all information must be transformed into these components through encoding.

Digitalization's long-term phenomenon is guiding all areas of understanding towards developing intricate models that involve elements and information, which can be interpreted by machines through coded language. Lovelace's acumen necessitates the implementation of "formalization," a concept that is reinforced by Jean-Claude Gardin (Gardin 1960). This serves as the bedrock for the true conceptualization of "digitality," where knowledge is predicated on a language that deconstructs meanings into increasingly discrete forms.

Gardin's essential was the 'translation' of text and archaeological finds into a data system, upon which, based on correct instructions, computers could bring their processing power to bear, to restore the complexity of archaeological information (Gardin 1959; 1962; 1989; 1991; Moscati 2013).

However, what information to translate? What to tell the machine, given that humanistic information cannot be entirely mathematized?

Digitality, as a historical phenomenon, entails the emergence of complexities that bring, in a more incisive way, reflection on the historical relationship between text and computable symbolism, which is quite different from computation, which finalizes the former (text) into the mechanization of reading.

The message that passes between two entities implies that there must be the same intellectual and rational conditions between them. In the case of historical sources, scholars, despite having the same intellectual abilities, do not always have the ontology of the subject who produced the document, but this does not mean that they will not be able to grasp the meanings of every grapheme. The computer, however —and therefore computation— only manages relations between symbols based on instructions, without being able to consider/infer meanings, which are different from signs, and therefore not computable — as Sebeok asserts, "sign[s] without either similarity or contiguity, but only with a conventional link between its signifier and its denotata" (Sebeok 1986).

What computers can do is a non-content-based "reasoning", thus simply numerical; and formalized.

On this concept, Jean-Claude Gardin begins his reflection, which will lead him to recognize in "encoding" the foundational role of humanistic methodological possibilities and to be the key to the New Computer Archaeology.

He lived in the fervent times of Turing's Machine; he appreciated its theorization, precisely because it was a process that did not detach from digitality, but rather brought the latter into a broader dimension, which would have empowered research, through the possibility of preserving archaeological knowledge on machine-readable media, which would have, on one hand, provided instructions, and on the other hand, would have had the task of organizing and storing.

This is the spirit that guided Gardin towards the foundation, in 1957, of the Centre Mécanographique de Documentation Archéologique. The centre represents the mature fruit of a research program related to a mission of the CNRS, in 1955, at the Institut Français d'Archéologie de Beyrouth, with the aim of creating a public archive of punched cards, for the cataloguing and development of a faster methodology in data retrieval, sorting, classification, and research.

Was this the Bushian memex (Bush 1945)? Probably, yes —or, at least, it is difficult to think that those decades were not carriers of reflections on the characteristics of computers and mainframes. However, the sense of a human process of computerization remains strong for the emergence of knowledge that must be a substance unit. Gardin's research was based on the "uncover the mental processes" at work in archaeological reasoning, in order to make them susceptible to Turing-like mechanical manipulation. The new method was based, indeed, on a coding workflow that ensured the segmentation and description of the minimal components of each object. The process identifies relationships between those components, through a formalized description, and then transcript them on punched cards. The latter, thanks to an automatic selection device, could make it possible to create virtual catalogues (Gardin 1960) and classifications, which could be used at different moments of archaeological research.

Therefore, encoding is the foundation of digitality. It is the possibility of decoding into processable terms that led the poet Josephine Miles to the project of "concordances" on Wordsworth's language (Miles 1942b; 1942a; 1946) —which represents decisive methodological evidence for the birth of a knowledge system that computes literary information (Spina in printing). This experience, after all, in the decade, and even in the subsequent years, will lead Miles to expand her project, aiming at the phrasal forms of poetry from the 1640s, 1740s, and 1840s (Miles 1948; 1950), attracting the attention of the Electrical Engineering Department of Berkeley, which invited her to direct a project on concordances in the works of John Dryden. However, the new project's time, however, through the use of a computer and punched cards. The result of this work is *The Continuity of Poetic Language: Studies in English Poetry from the 1540s to the 1940s (Miles 1951)*.

The age of concordances in the Computer Science field? Definitely; however, what we would like to point out is the real attention to what is at the foundation of "everything", namely, calculation, processability, and computing.

Punched cards, encoding, and mathematization remain central in Vannevar Bush's Memex, a tool that would have had the capacity to store a vast amount of information and make it possible to consult it at any time. The Memex aims to let scholars store documentation in a mechanized manner, and to become —as Murray Lawson states, while describing punched cards, adopting Bush's insights— a necessary tool precisely for historical research.

> «[T]he growing realization of the inadequacies of the conventional tools of research has forced upon historians a consideration of the equipment and techniques in use in the business world—a somewhat belated discovery on the part of historians, inasmuch as more than half a century ago, Hubert Howe Bancroft, the successful business man turned historian, advocated "the desirability of applying business methods [...] to historical [...] research". [...] In this paper an attempt will be made to describe two techniques in common use in the business world as well as to suggest some of the great advantages to be obtained from their application to the problems of historical research. The first is the marginally punched card system called Keysort and marketed in this country by the McBee Company. The second is the automatic punched card sorting and tabulating system developed by the International Business Machines Corporation and by Remington Rand Incorporated» (Lawson 1948).

We have to point out that Lawson's analysis does not focus on 'encoding' as a perspective for historical research, in a Turingian sense. Nonetheless, while looking at punched cards as opportunities to "open up vistas hitherto undreamt", he makes the latter the best tool for the "organization" of that bibliography and historiography which must be useful to historians and their projects, allowing for categorization of historical works, which can quickly emerge from computers, to meet the scholars' need to understand the state of the art and the organization of research results, for their publication.

However, according to Lawson, the only perspective of real application of the cards in historical research is given by "correlational" studies, thus for quantitative approaches. The possibility of organizing data "materially" allows scholars to reason about their historical problems, without any kind of "reductions" or selections among the information —often 'subjective'— from the materials to be analyzed, which, if on one hand is a necessary step towards easier control, on the other hand, forces scholars into a research project of "less ambitious proportions".

Therefore, in the discourse, the Lovelacean —then Turingian— perspective does not emerge, and Lawson makes "encoding" the «process of memorization on punched cards» aimed at organizing materials, leaving, in a certain way, to the scholar the task of analysis. Today, automatic notation tools, such as Zotero, Endnote, and Mendeley, precisely respond to the Lawsonian need to access a bibliographic database that can encompass existing historical production, in librarianship terms, capable of ensuring classification, accessibility, organization, uniqueness of entries, referencing, and publication.

Instead, it will be in the quantitative perspective (McCloskey 1978) that "encoding" will try to regain its uniqueness as a linguistic structure, in a perspective that will be at the basis of "social mathematics" (Braudel 2003) and Cliometrics (Fenoaltea 2019; Lyons 2007; McCloskey 1978). The latter, by the way, are not considered historical research, which will maintain —although Malthus (Cossa 1895; Malthus 2012), Ricardo (Ricardo 1887; 1891), and Le Play (Le Play 1875), have shown the need to "digitize/count" the data of a historical phenomenon (population, governability, family)— the paradigm, between criticism and openings (Chaunu 1964; Darcy and Rohrs 1995; Dollar and Jensen 1974; Erickson 1997; Furet 1981), that the object of that research field, which is "mankind", can never be mathematized, and therefore coded.

Subsequently, digitality disconnects from the conceptualization of "encoding", with the advent of GPUs (Graphic Cards), which allowed computer users to work without any knowledge of programming languages and coding, resulting, in digitality (and "digitization"), in the acquisition of a meaning unrelated to the whole, not Turingian: "manifestation". The process of digitization becomes synonymous with "photographing and displaying online" (Spina 2023a; 2022b; 2023b), through websites. Digitality becomes a manifestation of meaning that has nothing to do with encoding information —at least, this is what has characterized the last decades of the historical phenomenon. The advent of artificial intelligence platforms, instead, seems to be bringing the concept back to its original significant dimension. Digitization processes and projects are realigning with the needs of the Turing Machine to process and process. The TEI, metadata, tagging, and everything that underlies the construction of digital documents (digital editions), live from the idea and purpose of building accessible data heritages, both man and machine. The digital turn aims to create an informatic infrastructure, which, on one hand, is a dimension of existence of a new dialogical complexity, and on the other hand, aims at compiling an "ontologies' ontology" which is a connective structure of meaning, on which computer systems can build further knowledge, through computational analyses that, today, seem to have "convinced" historians in the possibility offered by tools and algorithms to assist them in analyzing past events.

## 5. Conclusion

History research itself is encoding, and Archives represent the first storing dimensions, where every information was codified to solve the aim to preserve in order to "recodify and create new knowledge." Historians and archivists codify and «process» all pieces of information stored on "archival sources (akin to cards)" to arrive at an output (the publication), which is the narration/reconstruction of the past. Ergo, Historiography is the result of the encoding process implemented by historians. For the same reason, we can point out that what we know about the world and our lives is provided by the coding process, which underlies digitality.

Thus, we can consider digitalization as a long-term phenomenon and an approach that underlies and has always underlies, the methodology of History, which is a general and generalizing act of codification, never closed or binding.

Thanks to Blaise Pascal, Gottfried Wilhelm Leibniz, Ada Lovelace, Charles Babbage, George Boole, Claude Elwood Shannon, and Alan Turing(Turing 1937; 1950), from the nineteenth century onwards, we witnessed the upgrade of theory about the possibility of communicating with machines, via a new concept of the "input phase", which required the use of a specific medium.

From that moment on, digitalization has had to grapple with two aspects of the human-machine relationship. On one hand, the need for encoding, and on the other hand, the development of increasingly high-performing storage mediums, in terms of data acquisition by computers.

Nowadays, the "acquisition phase" that characterizes the digital turn is still based on what we are going through, the collection of everything we own, in terms of physical heritage, to decode them into machine-readable code, accompanies humans and their desire to understand, first and foremost, themselves. The "acquisition phase" is the creation of the Big Archive of digitality, based on the Turing Machine's reasonings which require "discrete data", formalized in a language "apt to formulate an algorithm" (Orlandi 2021), and to create a new cognitive entity that needs to explain History also as the history of the digital ecological niche.

What Digital History needs, therefore, is a careful and deep reflection on computation as a long-term phenomenon, also to bridge the gap between rationality and intuition, through tools that will increasingly perfect themselves in their ability to grasp the importance of an assumption, aware that "deciding what it means" is still not the prerogative of the computer today.

Nevertheless, digitization is re-ontologizing the world, and encoding, while keeping humans at the centre of language, is making Computer Science the linguistic common denominator of the exchange structure, in which we also place computers, which, through long training phases, will be increasingly close to the possibility of providing meaningful interpretations for themselves and the scientific community; therefore, "reasoning" about the historical problem. For this reason, the historians' community must position itself as the unique matrix of the foundation of that ontology which will be the necessary mechanism of the DEN and of an AI platform that progressively enhances its algorithms.

Thus, it is necessary to return to the understanding of the "aim" of the Turing Machine, of its "internal" mechanism, and of the possibility/necessity of constructing a knowledge that can be remodelled in its linguistics, in its meanings, in its manifestations, with the aim of making the physical and digital world understandable to each other.

Historical Science and historians must therefore assume the role of the keystone of that ontology/dimension/system of coexistence of analogue and digital meanings. Historians are "Homo-Loggatus", but with the clear task of creating the Techo-biocenosis (Spina 2023c). The latter is the balance of the interplay systems, through a linguistic framework that allows for a narrative of Knowledge understandable to the machine, which is part of the cenosis. Therefore, the Turing machine became an "agent" that brings its possibilities into the search for that Hegelian historiographical completeness, «which must encompass all discourses... which are the axioms of the time» (Hegel 2014).

**References**

Babbage, Charles. 1989. *The Works of Charles Babbage: Mathematical Papers*. New York University Press.

———. 2010. "Note on the Application of Machinery to the Computation of Astronomical and Mathematical Tables." In *Babbage's Calculating Engines: Being a Collection of Papers Relating to Them; Their History and Construction*, edited by Henry P. Babbage, 211–211. Cambridge Library Collection - Mathematics. Cambridge: Cambridge University Press. https://doi.org/10.1017/CBO9780511694721.006.

Beyer, Kurt W. 2012. *Grace Hopper and the Invention of the Information Age*. MIT Press.

Bianchi, Claudia. 2003. *La filosofia di Gottlob Frege*. FrancoAngeli.

Braudel, Fernand. 2003. *Scritti sulla storia*. Milano: Bompiani. http://www.bompiani.it/catalogo/scritti-sulla-storia-9788845254123.

———. 2015. *Storia, misura del mondo*. Il Mulino.

Braudel, Fernand, and Alfredo Salsano. 1974. *La storia e le altre scienze sociali*. Roma: Laterza.

Bromley, Allan. 1991. *The Babbage Papers in the Science Museum Library: A Cross-Referenced List*. Science Museum.

Bush, Vannevar. 1945. "As We May Think." *The Atlantic*, 1945. https://www.theatlantic.com/magazine/archive/1945/07/as-we-may-think/303881/.

Casalegno, Daniele. 2013. *Uomini e computer. Storia delle macchine che hanno cambiato il mondo*. Milano: Hoepli.

Chaunu, Pierre. 1964. *Histoire quantitative ou histoire serielle*. Paris: Armand Colin.

Chomsky, Noam. 1975. *Reflection on language*. New York, N.Y: Pantheon.

Collier, Bruce, and James MacLachlan. 2000. *Charles Babbage: And the Engines of Perfection*. Oxford University Press, USA.

Corry, L. 2013. *David Hilbert and the Axiomatization of Physics (1898–1918): From Grundlagen Der Geometrie to Grundlagen Der Physik*. Springer Science & Business Media.

Cossa, Emilio. 1895. *Il principio di popolazione di Tomaso Roberto Malthus: saggio di economia sociale*. Libreria Treves di Pietro Virano.

Darcy, R, and Richard C Rohrs. 1995. *A Guide to Quantitative History*. Westport, Conn.: Praeger.

Dasgupta, Subrata. 2014. *It Began with Babbage: The Genesis of Computer Science*. Oxford University Press.

Dollar, Charles M, and Richard J Jensen. 1974. *Historian's Guide to Statistics; Quantitative Analysis and Historical Research*. Huntington, N.Y.: R.E. Krieger Pub. Co.

Dunbar, Robin Ian MacDonald. 1998. *Grooming, Gossip, and the Evolution of Language*. First Paperback edition. Cambridge, Mass: Harvard University Press.

Duranti, Alessandro. 2005. *Antropologia del linguaggio*. Meltemi Editore srl.

Erickson, Bonnie H. 1997. "Social Networks and History: A Review Essay." *Historical Methods: A Journal of Quantitative and Interdisciplinary History* 30 (3): 149–57. https://doi.org/10.1080/01615449709601182.

Fenoaltea, Stefano. 2019. "Spleen : The Failures of the Cliometric School." *Annali Fondazione Einaudi* 53 (2): 5–24.

Floridi, Luciano. 2012. *La rivoluzione dell'informazione*. Torino: Codice. http://trentino.medialibrary.it/media/scheda.aspx?idm=20146&idtm=30015&page=0&preferiti=.

———. 2016. *The 4th Revolution, How the Infosphere Is Reshaping Human Reality*. Oxford; New York: Oxford University Press.

———. 2020. *Pensare l'infosfera. La filosofia come design concettuale*. Raffaello Cortina Editore.

Frege, Gottlob. 1980. *The Foundations of Arithmetic: A Logico-Mathematical Enquiry Into the Concept of Number*. Northwestern University Press.

———. 2007. *Senso, funzione e concetto. Scritti filosofici 1891-1897*. Laterza.

———. 2021. *L'alfabeto del pensiero*. LIT EDIZIONI.

———. 2023. *On Sense and Reference*. DigiCat.

Furet, François. 1981. "Il Quantitativo in Storia." In *Fare Storia. Temi e Metodi Della Nuova Storiografia*, edited by Jacques Le Goff and Pierre Nora. Torino: Einaudi.

Gardin, Jean-Claude. 1959. "Four Codes for the Description of Artefacts. An Essay in Archaeological Technique and Theory." *American Anthropologist* 60: 335–57.

———. 1960. "Les applications de la mécanographie dans la documentation archéologique." *Bulletin des Bibliothèques de France* 5: 1–16.

———. 1962. "De l'archéologie à l'information Automatique." *Bulletin Euratom* 4: 25–29.

———. 1989. "Artificial Intelligence and the Future of Semiotics: An Archaeological Perspective" 77 (1–3): 5–26. https://doi.org/10.1515/semi.1989.77.1-3.5.

———. 1991. *Le calcul et la raison. Essais sur la formalisation du discours savant*. Paris: Éditions de l'École des Hautes Études en Sciences Sociales.

Gödel, Kurt. 1940. *The Consistency of the Axiom of Choice and of the Generalized Continuum-Hypothesis with the Axioms of Set Theory*. Princeton University Press.

———. 2012. *On Formally Undecidable Propositions of Principia Mathematica and Related Systems*. Courier Corporation.

Godfrey-Smith, Peter. 2000. "Niche Construction in Biological and Philosophical Theories." *Behavioral and Brain Sciences* 23 (1): 153–54. https://doi.org/10.1017/s0140525x00312419.

Haarmann, Harald. 2021. *Storia universale delle lingue. Dalle origini all'era digitale*. Translated by Claudia Acher Marinelli. 3° edizione. Bollati Boringhieri.

Hartog, François. 2015. *Regimes of Historicity: Presentism and Experiences of Time*. Translated by Saskia Brown. Columbia University Press.

Hegel, Georg Wilhelm Friedrich. 2014. *Filosofia della storia universale. Secondo il corso tenuto nel semestre invernale 1822-23*. Torino: Einaudi. https://www.einaudi.it/catalogo-libri/senza-materia/filosofia-della-storia-universale-georg-wilhelm-friedrich-hegel-9788858415535/.

Hilbert, David. 1893. "Ueber die vollen Invariantensysteme." *Mathematische Annalen* 42 (3): 313–73. https://doi.org/10.1007/BF01444162.

———. 1972. *Grundzüge der Theoretischen Logik*. 6. Aufl. 1959 edizione. Berlin: Springer.

———. 2012. *Grundlagen der Mathematik*. Springer.

———. 2015. *The Foundations of Geometry*. Read Books Ltd.

———. 2021. *Mathematical Problems*. Good Press.

Hilbert, David, and Stephan Cohn-Vossen. 1999. *Geometry and the Imagination*. American Mathematical Soc.

Hyman, Anthony. 1985. *Charles Babbage: Pioneer of the Computer*. Princeton University Press.

Iannucci, Giulia. 2020. "Naturalizzare l'individuo: Menti, Soggettività e Nicchie Ecologiche." Roma: Università degli Studi di Roma "La Sapienza." https://iris.uniroma1.it/handle/11573/1360660.

Johansson, Sverker. 2023. *L'alba del linguaggio. Come e perché i Sapiens hanno iniziato a parlare*. Ponte alle Grazie.

Laland, Kevin N., and Michael J. O'Brien. 2011. "Cultural Niche Construction: An Introduction." *Biological Theory* 6 (3): 191–202. https://doi.org/10.1007/s13752-012-0026-6.

Laland, Kevin N., F. John Odling-Smee, and Marcus W. Feldman. 2000a. "Group Selection: A Niche Construction Perspective." *Journal of Consciousness Studies* 7 (1–2): 1–2.

Laland, Kevin N., John Odling-Smee, and Marcus W. Feldman. 2000b. "Niche Construction, Biological Evolution, and Cultural Change." *Behavioral and Brain Sciences* 23 (1): 131–46. https://doi.org/10.1017/s0140525x00002417.

———. 2005. "On the Breadth and Significance of Niche Construction: A Reply to Griffiths, Okasha and Sterelny." *Biology and Philosophy* 20 (1): 37–55. https://doi.org/10.1007/s10539-004-6834-8.

Latina, Massimiliano. 2013. "Nicchie Ecologiche e Nicchia Ontologica. Una Riflessione Tra Le Teorie Della Niche Construction e La Lichtung Di Peter Sloterdijk." S&F_scienzaefilosofia.It.

Lawson, Murray G. 1948. "The Machine Age in Historical Research." *The American Archivist* 11 (2): 141–49.

Lazard, Emmanuel, and Pierre Mounier-Kuhn. 2016. *Histoire illustrée de l'informatique: Histoire illustrée de l'informatique*. Illustrated édition. Les Ulis: EDP SCIENCES.

Le Play, Frédéric. 1875. *L'Organisation de La Famille, Selon Le Vrai Modèle Signalé Par l'histoire de Toutes Les Races et de Tous Les Temps, Par M. F. Le Play,... Avec Trois Appendices, Par MM. É. Cheysson, F. Le Play et C. Jannet. 2e Édition...* Paris: Dentu Libraire. https://gallica.bnf.fr/ark:/12148/bpt6k6535021z.

Le Roy Ladurie, Emmanuel. 1968. "La Fin Des Érudits. L'historien de Demain Sera Programmeur Ou Ne Sera Pas." *Nouvel Observateur*, mai 1968.

Liskov, B., and John Guttag. 1986. *Abstraction and Specification in Program Development*. MIT Press.

Lotman, Jurij. 1984. "Bachtin – sein Erbe und aktuelle Probleme der Semiotik." In *Roman und Gesellschaft: Internationales Michail-Bachtin-Colloquium*, edited by Hans-Günter Hilbert. Jena: Friedrich-Schiller-Universität.

Lovelace, Ada. 1843. "Sketch of the Analytical Engine Invented by Charles Babbage." *Scientific Memoirs, Selected from the Transactions of Foreign Academies of Science and Learned Societies*, no. 3: 666–731.

Lyons, J.S. 2007. "Introduction: Economic History and Cliometrics." In *Reflections on the Cliometrics Revolution: Conversations with Economic Historians*, by P. Cain, S.H. Williamson, and J.S. Lyons, 1–42. London: Routledge.

Malthus, T. R. 2012. *An Essay on the Principle of Population*. Courier Corporation.

McCloskey, Donald N. 1978. "The Achievements of the Cliometric School." *The Journal of Economic History* 38 (1): 13–28.

Menabrea, Luigi Federico. 1842. "Notions Sur La Machine Analitique de M. Charles Babbage." *Bibliothèque Universelle de Genève*, 1–28.

Miles, Josephine. 1942a. *Pathetic Fallacy in the 19th Century: A Study of a Changing Relation Between Object and Emotion*. University of California Press.

———. 1942b. *Wordsworth and the Vocabulary of Emotion*. University of California Press.

———. 1946. *Major Adjectives in English Poetry from Wyatt to Auden*. University of California Press.

———. 1948. *The Primary Language of Poetry in the 1640's*. University of California press.

———. 1950. *The Primary Language of Poetry in the 1740's and 1840's*. University of California Press.

———. 1951. *The Continuity of Poetic Language: Studies in English Poetry from the 1540's to the 1940's*.

Moscati, P. 2013. "Jean-Claude Gardin (Parigi 1925-2013). Dalla meccanografica all'informatica archeologica." *Archeologia e Calcolatori* 24: 7–24. https://doi.org/10/1/01_Moscati.pdf.

Nadel, George H. 1964. "Philosophy of History Before Historicism." *History and Theory* 3 (3): 291–315. https://doi.org/10.2307/2504234.

Nahin, Paul J. 2017. *The Logician and the Engineer: How George Boole and Claude Shannon Created the Information Age*. Princeton University Press.

Odling-Smee, John, and J. Scott Turner. 2011. "Niche Construction Theory and Human Architecture." *Biological Theory* 6 (3): 283–89. https://doi.org/10.1007/s13752-012-0029-3.

Orlandi, Tito. 2021. "Reflections on the Development of Digital Humanities." *Digital Scholarship in the Humanities* 36 (Supplement 2): ii222–29. https://doi.org/10.1093/llc/fqaa048.

Ricardo, David. 1887. *Letters of David Ricardo to Thomas Robert Malthus, 1810-1823*. Clarendon Press.

———. 1891. *Principles of Political Economy and Taxation*. G. Bell and sons.

Rodriguez-Consuegra, Francisco. 1995. *Kurt Gödel: Unpublished Philosophical Essays*. Springer Science & Business Media.

Roselló, Joan. 2019. *Hilbert, Göttingen and the Development of Modern Mathematics*. Cambridge Scholars Publishing.

Saussure, Ferdinand de. 1879. *Mémoire Sur Le Système Primitif Des Voyelles Dans Les Langues Indo-Européennes*. https://gallica.bnf.fr/ark:/12148/bpt6k729200.

———. 2014. *De l'emploi Du Génitif Absolu En Sanscrit: Thèse Pour Le Doctorat Présentée à La Faculté de Philosophie de l'Université de Leipzig*. Cambridge Library Collection - Linguistics. Cambridge: Cambridge University Press. https://doi.org/10.1017/CBO9781107589889.

Sebeok, Thomas A. 1986. *Encyclopedic Dictionary of Semiotics*. Berlin, Boston: De Gruyter Mouton. https://www.degruyter.com/isbn/9783112321294.

Shannon, Claude Elwood. 1948. "A Mathematical Theory of Communication." *The Bell System Technical Journal* 27 (3): 379–423. https://doi.org/10.1002/j.1538-7305.1948.tb01338.x.

———. 1993. *Claude E. Shannon: Collected Papers*. Wiley.

Shannon, Claude Elwood, and Warren Weaver. 1998. *The Mathematical Theory of Communication*. University of Illinois Press.

Slater, Robert. 1989. *Portraits in Silicon*. MIT Press.

Spina, Salvatore. 2022a. *Digital History. Metodologie Informatiche per La Ricerca Storica*. Napoli: Edizioni Scientifiche Italiane.

———. 2022b. "Historical Network Analysis & Htr Tool. Per un approccio storico metodologico digitale all'archivio Biscari di Catania." *Umanistica Digitale*, no. 14: 163–81. https://doi.org/10.6092/issn.2532-8816/15159.

———. 2023a. "Artificial Intelligence in Archival and Historical Scholarship Workflow: HTR and ChatGPT." *Umanistica Digitale*. https://doi.org/10.48550/arXiv.2308.02044.

———. 2023b. "Biscari Epistolography. From Archive to the Website." *DigItalia* 2.

———. 2023c. "Homo-Loggatus. The Anthropological Condition of Historians in the Digital World." *Journal of Mathematical Techniques and Computational Mathematics* 2 (10): 431–37.

———. 2023d. "Homo-loggatus? Come stare dentro la nicchia ecologica digitale." Billet. *Informatica Umanistica e Cultura Digitale: il blog dell'AIUCD* (blog). July 19, 2023. https://infouma.hypotheses.org/2224.

———. in printing. "Per Una Storia Della Storiografia Della Digital History (DHy)." *Nuova Rivista Storica*.

Steen, Wim J. van der. 2000. "Niche Construction: A Pervasive Force in Evolution?" *Behavioral and Brain Sciences* 23 (1): 162–63. https://doi.org/10.1017/s0140525x00422417.

Swade, Doron. 2002. *The Difference Engine: Charles Babbage and the Quest to Build the First Computer*. Penguin Books.

Todorov, T., and G. L. Bravo. 2003. *I formalisti russi. Teoria della letteratura e metodo critico*. Einaudi.

Turing, Alan Mathison. 1937. "On Computable Numbers, with an Application to the Entscheidungsproblem." *Proceedings of the London Mathematical Society* s2-42 (1): 230–65.

———. 1950. "Computing Machinery and Intelligence." *Mind* LIX (236): 433–60. https://doi.org/10.1093/mind/LIX.236.433.

Uller, Tobias, and Heikki Helanterä. 2019. "Niche Construction and Conceptual Change in Evolutionary Biology." *British Journal for the Philosophy of Science* 70 (2): 351–75. https://doi.org/10.1093/bjps/axx050.

Wallmark, Laurie. 2017. *Grace Hopper: Queen of Computer Code.* Sterling Children's Books.

Whitehead, Alfred North, Bertrand Russell, and Kurt Gödel. 1986. *Principia mathematica: Vorwort und Einleitungen.* Suhrkamp.

Wiener, Norbert. 1953. *La cibernetica.* Milano: Bompiani.

Wirth, Niklaus. 2008. *Algorithms + Data Structures = Programs.* New Delhi: Prentice-Hall of India.